\documentclass[conference]{IEEEtran}
\IEEEoverridecommandlockouts
\usepackage{cite}
\usepackage{amsmath,amssymb,amsfonts}
\usepackage{algorithmic}
\usepackage{graphicx}
\usepackage{textcomp}
\usepackage{xcolor}
\usepackage{amsmath}
\usepackage{graphicx}
\usepackage{array}
\usepackage{tabularx}
\usepackage{booktabs}

\newcommand{\tool}[0]{HumanEval\textunderscore T }

\newcommand{\TODO}[2][1]{%
    \marginnote{\color{red}\textbf{TODO}\ifnum#1>1\space(!)\fi}%
    {\color{red!70!black}\hl{#2}}%
}

\def\BibTeX{{\rm B\kern-.05em{\sc i\kern-.025em b}\kern-.08em
    T\kern-.1667em\lower.7ex\hbox{E}\kern-.125emX}}
\begin{document}

\title{Addressing Data Leakage in HumanEval Using Combinatorial Test Design\\\thanks{This research was supported by the Natural Sciences and Engineering Research Council of Canada (NSERC), grant 2018-06588, and by MITACS.}
}

\author{\IEEEauthorblockN{Jeremy S. Bradbury}
\IEEEauthorblockA{\textit{Ontario Tech University} \\
Oshawa, ON, Canada\\
jeremy.bradbury@ontariotechu.ca}
\and
\IEEEauthorblockN{Riddhi More}
\IEEEauthorblockA{\IEEEauthorblockA{\textit{Ontario Tech University} \\
Oshawa, ON, Canada\\
riddhi.more1@ontariotechu.net}
}}

\maketitle

\begin{abstract}
The use of large language models (LLMs) is widespread across many domains, including Software Engineering, where they have been used to automate tasks such as program generation and test classification. As LLM-based methods continue to evolve, it is important that we define clear and robust methods that fairly evaluate performance. Benchmarks are a common approach to assess LLMs with respect to their ability to solve problem-specific tasks as well as assess different versions of an LLM to solve tasks over time. For example, the HumanEval benchmark is composed of 164 hand-crafted tasks and has become an important tool in assessing LLM-based program generation. However, a major barrier to a fair evaluation of LLMs using benchmarks like HumanEval is data contamination resulting from data leakage of benchmark tasks and solutions into the training data set. This barrier is compounded by the black-box nature of LLM training data which makes it difficult to even know if data leakage has occurred. To address the data leakage problem, we propose a new benchmark construction method where a benchmark is composed of template tasks that can be instantiated into new concrete tasks using combinatorial test design. Concrete tasks for the same template task must be different enough that data leakage has minimal impact and similar enough that the tasks are interchangeable with respect to performance evaluation. To assess our benchmark construction method, we propose \tool, an alternative benchmark to HumanEval that was constructed using template tasks and combinatorial test design.
\end{abstract}


\section{Introduction}

The proliferation of LLMs to address Software Engineering problems and tasks has been well reported in the literature~\cite{FGH+23, WHC+24, HZL+24}. One of the earliest problems to be addressed by LLMs was program generation where a natural language description is provided as a prompt to an LLM and a source code solution is generated. Other problems that have been addressed include test classification, clone detection and program repair.

As more LLM-based techniques are developed it is important to have data and methods to compare the efficacy of these techniques as well as to measure progress as LLMs are further trained and evolved. Benchmarking is a common method for evaluating the success of LLMs to solve specific problems or tasks~\cite{CWW+24}. Benchmarking has also been used successfully in software engineering to assess LLM-based techniques~\cite{HJHC24}. Example LLM benchmarks have been developed for code generation~\cite{CTJ+21, YSR+24}, software modeling~\cite{CBT24}, cs concepts in education~\cite{ARMX24}, and program repair~\cite{WLZL24}. 

While benchmarks have been utilized successfully they are not without their criticisms and limitations. For example, LLM evaluations utilizing benchmarks may exhibit bias and unfair assessment which needs to be understood and measured~\cite{DSK+21, LBL+22}. In addition to potential bias and fairness considerations, benchmarks also have limitations over their lifetime due to data leakage and data contamination~\cite{ZZC+23, SCG+23, LF24}. 
LLMs that have experienced data leakage and contamination during training will likely exhibit inflated benchmark evaluation scores which can misrepresent their ability to address the underlying Software Engineering task as well as misrepresent their performance when compared to other LLMs that have not experienced data leakage and contamination with the benchmark data. 

To address data leakage and evaluation fairness, we propose a new benchmark construction method where a benchmark is composed of template tasks that can be instantiated into new concrete tasks using combinatorial test design~\cite{TL02,CU10}. Concrete tasks for the same template task must be different enough that data leakage has minimal impact and similar enough that the tasks are interchangeable with respect to evaluation. To demonstrate our approach we have developed \tool, a variant of HumanEval, that was constructed using our proposed method. We have not yet developed a full \tool alternative to all of HumanEval's 164 tasks -- instead we have randomly sampled 10 of the HumanEval tasks and created \tool alternatives for this subset of the benchmark. Using the random subsets of \tool and HumanEval we answer the following research questions:

\begin{figure*}[t!]
\centering
\includegraphics[width=14cm]{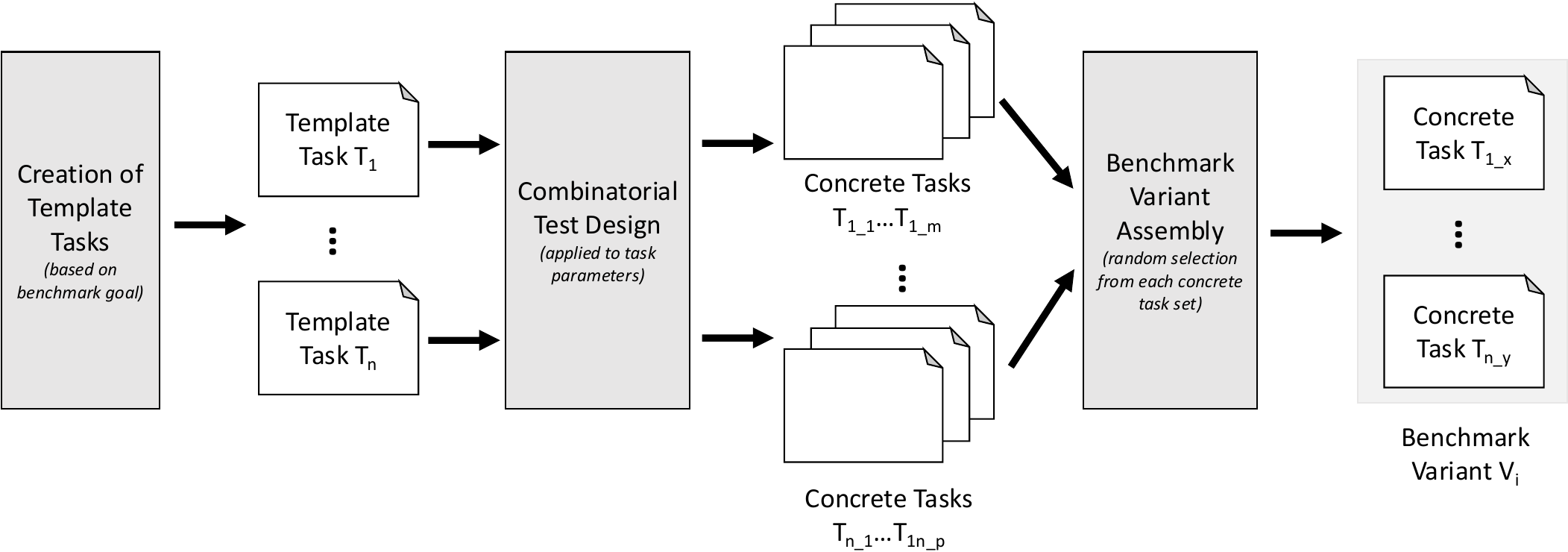}
\caption{Benchmark Construction Approach Using Combinatorial Test Design}
\label{fig:benchmark_process}
\end{figure*}

\begin{enumerate}
    \item Is there evidence of data leakage when comparing \tool with HumanEval?
    \item Do concrete versions of the same template task in \tool produce similar results?
\end{enumerate}



\section{Background}

\subsection{Benchmarking}

Benchmarks serve as standardized evaluation tools that enable systematic comparison of different technologies and approaches that solve the same problem or task~\cite{SEH03}. The emergence of Large Language Models (LLMs) has introduced new challenges in benchmark design and evaluation methodology that can impact fairness.

Data leakage represents a significant threat to LLM benchmark integrity. As documented by Sainz et al.~\cite{SCG+23}, contamination occurs when benchmark data appears in an LLM's training set, leading to artificially inflated performance metrics and potentially invalid scientific conclusions. This issue is particularly acute given the black-box nature of many LLMs and the difficulty in determining training data composition. 

For example, let's consider the HumanEval~\cite{CTJ+21} benchmark which is comprised of 164 hand-crafted programming tasks. HumanEval has become a standard benchmark for evaluating code generation capabilities, however, a recent study by Li and Flanigan~\cite{LF24} demonstrated the risks of data contamination. They found that task contamination significantly impacted model performance, with pre-training exposure to benchmark tasks resulting in misleadingly high scores.

\subsection{Combinatorial Testing}

In software testing, combinatorial test design enables efficient coverage of parameter spaces while minimizing test case count~\cite{CU10}. Previous work leveraged combinatorial test design in the construction of a concurrency bug benchmark as a way to ensure diversity while minimizing the instances of benchmark tasks~\cite{BSF+12}. More recent work has extended the use of combinatorial test design into LLM evaluation, where combinatorial testing is used to generate diverse test cases while preserving semantic consistency~\cite{GKL+23} . 

\section{Related Work}

To address the challenge of data leakage and fairness in Software Engineering LLM benchmarks, a number of best practices have been identified regarding the construction and maintenance of benchmarks. With respect to construction, it is important to ensure benchmark tasks do not originate from sources that are part of any LLM training data sets.The two most common strategies for ensuring this separation of benchmark data and LLM training data are to (1) hand-craft benchmark tasks (e.g., HumanEval) or (2) select benchmark tasks from private data sets that are not accessible for LLM training. With respect to benchmark operation and maintenance, it is important that benchmark data be actively excluded from future LLM training data sets. In cases where these benchmark construction and maintenance best practices are followed, the risk of data leakage is low. However, with the prevelence of black-box training in commerical LLMs, it is impossible to know if the benchmark data has been leaked as it relies on trust in LLM developers to abide by the best practice of excluding benchmark data from training. Even when efforts are made to exclude benchmark data sources from training data, leakage can still occur due to third-party training sources that duplicate, summarize or even mention a benchmark's data. 

Another approach to address benchmark fairness is to create dynamic or evolving benchmarks using agents~\cite{WLF+24, ZWZ+24}. The goal of these benchmarks is to evaluate emerging capabilities or evaluate multi-faceted capabilities by dynamically evolving the benchmark. This evolution has the side effect that benchmark tasks are not identical over time and thus are less likely to be impacted by data contamination. However, evolving data creates challenges when comparing LLMs against historic benchmark evaluations as LLMs are continuously evaluated against an evolving standard. Furthermore, removing  old benchmark tasks or modifying them creates a blind spot to possible model regression where future LLMs maybe unable to effectively solve previously evaluated tasks. 








\section{A Benchmark Construction Approach to Mitigate Data Leakage}

Recent work demonstrates that data leakage significantly impacts LLM evaluation reliability~\cite{SCG+23,ZZC+23}. We propose a novel benchmark construction methodology based on template tasks and combinatorial testing to mitigate data leakage and provide a fair evaluation. Our approach involves three key phases: template task creation, combinatorial test design and benchmark variant assembly (see Fig.~\ref{fig:benchmark_process}). To demonstrate our approach we will use HumanEval problem \#1 as an example. This problem tests proximity comparison in numeric arrays. The original problem states\footnote{HumanEval also includes example input and output pairs with each problem however we have excluded these example pairs from our analysis.}:

\begin{small}
\begin{quote}
\texttt{
Given a list of numbers and a threshold value, determine if any two values are closer than the threshold.}
\end{quote}
\end{small}

\subsection{Template Task Creation}

Rather than static benchmark tasks, we define template tasks that capture fundamental task properties while abstracting some implementation details. For example, we transform HumanEval Problem \#1 into the following template tasks: 

\begin{small}

\begin{quote}
\texttt{Given a list of \textbf{<input\_type>} and a \textbf{<threshold\_descriptor>}, check if any two \textbf{<value\_descriptor>} are closer than the given \textbf{<threshold\_descriptor>}.}

\texttt{Where:}
\begin{itemize}
\item \texttt{<input\_type>: numbers, float values, measurements}
\item \texttt{<threshold\_descriptor>: threshold, minimum distance, tolerance}
\item \texttt{<value\_descriptor>: values, elements, data points}
\end{itemize}
\end{quote}
\end{small}
\noindent In the above template task we first generalized the HumanEval problem description by replacing specific details with template variables that can systematically be replacing with the original values or alternative values. For example, in the above example the {\texttt{<input\_type>} template variable can be replaced with numbers, float values or measurements. It is important to ensure that template variables are not used to replace elements that impact the difficulty or purpose of the problem. In an effort to maintain 
semantic behavioural equivalence we have focused on varying input/output data types, variable naming and descriptions as well as problem context and framing.



\subsection{Combinatorial Task Design}

Once we have defined a set of template tasks, we employ combinatorial test design techniques~\cite{CU10} to generate a set of concrete task instances for each template task.  
The goal of this approach is to systematically create varied instances while maintaining semantic equivalence and comparable complexity across them. 

If we consider the above template task for HumanEval problem \#1 and use combinatorial test design, we can systematically generate a set of concrete tasks that include:

\begin{small}
\begin{enumerate}
\item \texttt{Given a list of numbers and a threshold, check if any two values are closer than the given threshold}
\item \texttt{Given a list of measurements and a minimum distance, check if any two data points are closer than given minimum distance}
\item \texttt{Given a list of float values and a tolerance, check if any two elements are closer than the given tolerance}
\end{enumerate}
\end{small}

\noindent Each variant maintains the same computational objective while presenting the problem differently. 
Each concrete variant also includes a manually created set of test cases that validate the same underlying computational skills, helping detect whether performance differences stem from data leakage rather than genuine problem-solving ability~\cite{CTJ+21}.

\subsection{Benchmark Variant Assembly}

Once we have produce a set of concrete tasks for each template task using combinatorial test design we then create concrete benchmark variants -- each of which contains a unique concrete task for each template task that is not used in other benchmark variants. The concrete tasks are selected randomly to compose each benchmark variant. 
The use of different benchmark variants reduces the impact of direct memorization from training data with a goal of maintaining evaluation fairness across different LLM evaluations. 

\section{Experimental Assessment of Benchmark Task Construction}

\subsection{Experimental Setup}

To assess the benefits of our combinatorial test design approach to benchmark construction, we conducted experiments using a 10 problem subset from HumanEval~\cite{CTJ+21}. Each problem was converted into a template task with combinatorial parameter variations. We evaluated the HumanEval subset and \tool variant benchmarks using four state-of-the-art LLMs: GPT-3.5, GPT-4o, Claude 3.5 Sonnet and Llama 3.1. The LLMs are listed in order of release date.

For each template task, we generated five concrete versions through combinatorial testing
~\cite{GKL+23} and assemble the concrete versions into five variant benchmarks. 
We measure the code generation performance using \textit{pass@1}
~\cite{CTJ+21}
, where each model gets 1 attempts to generate a correct code that is validated against a test suite. In order to obtain a more fine-grained analysis we report back the percentage of tests that pass and not just if the generated code is correct or incorrect\footnote{All results reported are the average of performing this analysis five times.}. 

\subsection{Experimental Results}

We present our findings with respect to 
our two research questions: evidence of data leakage in HumanEval and consistency across \tool template-task versions.

\begin{figure}[t!]
\centering
\includegraphics[width=\linewidth]{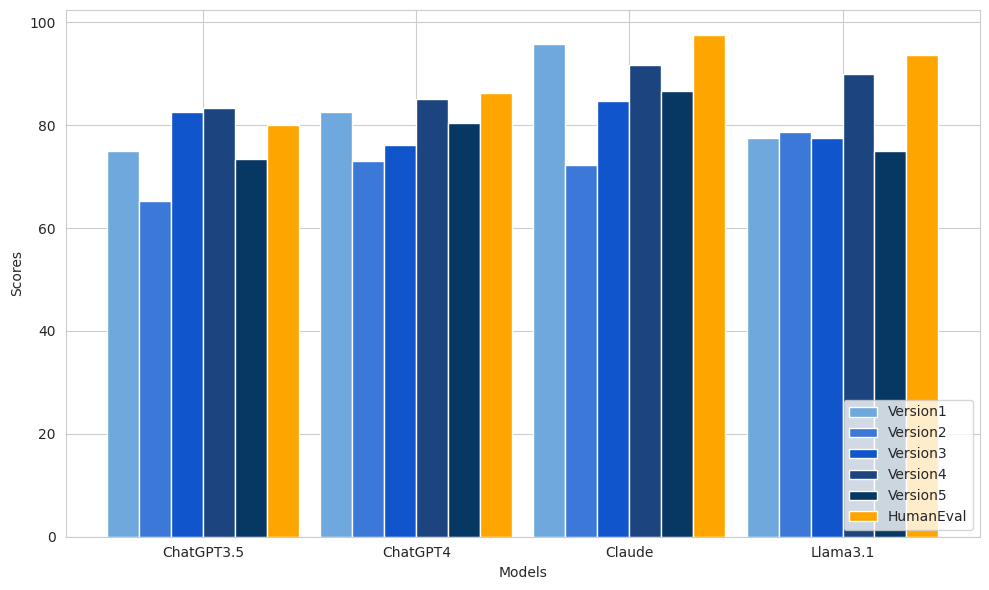}
\caption{Performance Distribution Across \tool variants (V1-V5) with HumanEval}
\label{fig:performance_distribution}
\end{figure}

\begin{table}[t!]
\centering
\small 
\setlength{\tabcolsep}{3pt} 
\renewcommand{\arraystretch}{1.2} 
\begin{tabular}{l|ccccccc}
\hline
\textbf{Model} & \textbf{V1} & \textbf{V2} & \textbf{V3} & \textbf{V4} & \textbf{V5} & \textbf{AVG} & \textbf{HE} \\ \hline
GPT3.5 & 75.0 & 65.3 & 82.6 & 83.3 & 73.4 & 76.7 & 80.0 \\ 
GPT4o & 82.5 & 73.1 & 76.2 & 85.0 & 80.5 & 79.4 & 86.2 \\ 
Claude 3.5 & 95.8 & 72.2 & 84.7 & 91.6 & 86.7 & 86.2 & 97.5 \\
Llama3.1 & 77.5 & 78.7 & 77.5 & 90.0 & 75.0 & 79.7& 93.7\\ \hline
\end{tabular}
\vspace{2 mm}
\caption{Model Performance Across \tool Benchmark Variants and HumanEval} 
\vspace{-3mm}
\footnotesize{\textbf{Note:} V1-V5 refer to the \tool variants, AVG is the average of V1-V5, and HE is HumanEval.}
\label{table:modelcomparison}
\end{table}

\subsubsection{RQ1: Evidence of Data Leakage}

To investigate potential data leakage in HumanEval, we compared model performance between the original HumanEval benchmark and our \tool variant benchmarks. Fig.~\ref{fig:performance_distribution} and Tab.~\ref{table:modelcomparison} illustrate these results when comparing the average percentage of tests passed per problem. Our analysis indicates potential data leakage when comparing the results of the \tool variants with HumanEval.

All models show notably higher performance on HumanEval compared to the \tool variants:
\begin{itemize}
\item GPT-3.5: 4.75\% drop (80\% $\rightarrow$ 75.25\%)
\item GPT-4o: 6.76\% drop (86.25\% $\rightarrow$ 79.49\%)
\item Claude: 11.27\% drop (97.5\% $\rightarrow$ 86.23\%)
\item Llama 3.1: 13.75\% drop (93.75\% $\rightarrow$ 79.75\%)
\end{itemize}

The consistent performance drop across all models indicates likely data leakage in the original benchmark. In three of the four model evaluations, HumanEval's results are identified an outlier, meaning their variation is more extreme than the expected (see Fig.~\ref{fig:version_scores}).

\subsubsection{RQ2: Consistency Across \tool Variants}

To evaluate whether our \tool variants produce consistent results, we analyzed the variation in performance as shown in Figure~\ref{fig:version_scores}. Recall, that consistency across \tool variants indicates that the concrete tasks generated via combinatorial test design are likely interchangeable as tasks across benchmark variants.

\begin{figure}[t!]
\centering
\includegraphics[width=\linewidth]{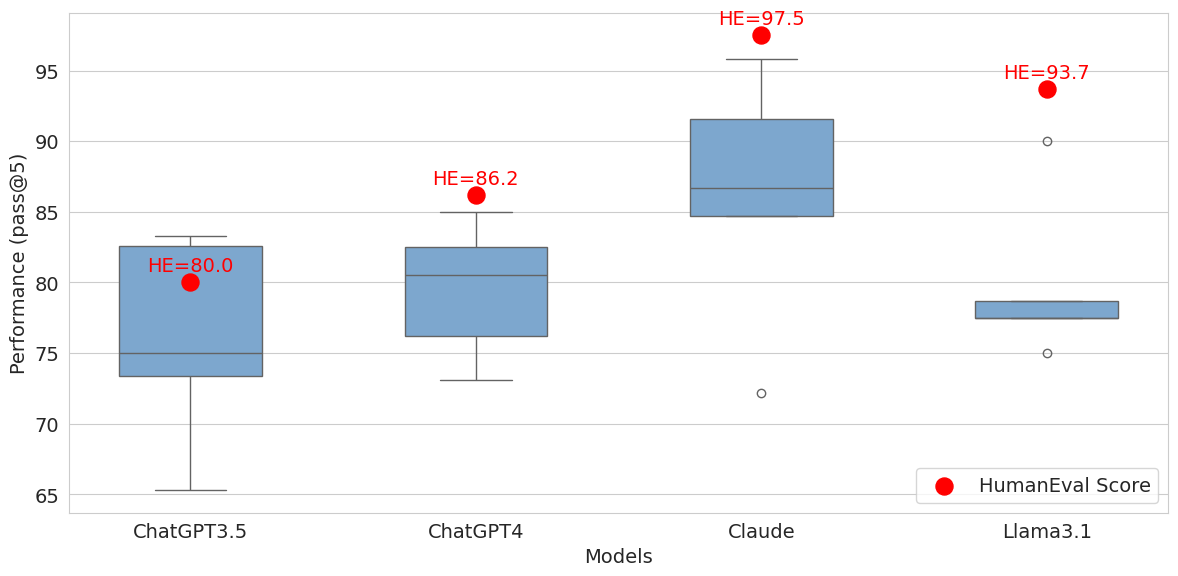}
\caption{Performance Comparison of Models with HumanEval scores }
\label{fig:version_scores}
\end{figure}

Analysis of performance across \tool variants reveals interesting results. GPT-3.5 shows moderate consistency ($\sigma = 7.6\%$) with variants performing a modest $4.75\%$ below HumanEval. GPT-4 demonstrates varied performance ($\sigma = 6.9\%$) across variants, with a notably higher $6.76\%$ gap from HumanEval. Claude shows the highest variation ($\sigma = 8.9\%$) in variant performance, with an $11.27\%$ difference from HumanEval. Llama 3.1 exhibits consistent performance across variants ($\sigma = 1.8\%$), yet shows the most significant drop of $13.75\%$ from the HumanEval baseline, indicating a trend of increasing performance gaps in newer models. 

\subsection{Discussion}

The experimental results while not conclusive reveal significant insights:

\begin{enumerate}
\item The systematic performance difference between HumanEval and \tool variants suggests static benchmarks have limitations and a relatively short period of usefulness for assessing models against historic benchmark results. While all models perform well on the original HumanEval tasks, their performance on the \tool variants, which were designed to be semantically comparable, varies consistently. This result aligns with recent findings about the evolution of model capabilities and the need for dynamic evaluation approaches~\cite{LF24}.

\item Our combinatorial test design approach to benchmark construction enables continuous benchmark evolution through semantically comparable task variations. While further analysis with a larger set of tasks is needed, we are optimistic that our approach can allow the creation of benchmark variants that preserve task complexity while facilitating fair model comparisons over time. If proven successful, this design supports longitudinal studies of model improvement while controlling for benchmark data leakage.
\end{enumerate}


\section{Conclusions and Future Work}

Our combinatorial test design approach to benchmark construction offers a systematic  method for creating benchmark variants with a goal of being robust to potential data leakage of earlier variants. 
Initial results from our evaluation of a subset of HumanEval tasks compared with five \tool benchmark variants demonstrates this method's potential for reducing the impact of data leakage in benchmark evaluation over time and in distinguishing between enhanced model capabilities versus performance improvements based on data leakage.

Future work needs to be done to fully evaluate our benchmark construction approach, first in the context of HumanEval and later in the context of other static LLM benchmarks. In the short-term we will focus on expanding \tool to encompass the complete set of 164 HumanEval tasks, while further refining our template creation process to ensure semantic comparability across concrete task versions. We also plan to investigate metrics for assessing consistency in the difficulty of concrete tasks generated from the same template tak. 
In the long-term we are also interested in applying our benchmark construction methodology to other benchmarks in other software engineering domains such as program repair and clone detection. 

\bibliographystyle{IEEEtran}
\bibliography{IEEEabrv,references}

\end{document}